# The necessity and power of random, under-sampled experiments in biology

**Brian Cleary[1] and Aviv Regev[2,3,4,5]**

[1]Broad Institute of MIT and Harvard, Cambridge, MA 02142, USA

[2]Department of Biology, Massachusetts Institute of Technology, Cambridge, MA 02139, USA

[3]Howard Hughes Medical Institute, Chevy Chase, MD, USA

[4]Klarman Cell Observatory, Broad Institute of MIT and Harvard

[5]Current address: Genentech, 1 DNA Way, South San Francisco, CA, USA

Correspondence should be addressed to bcleary@broadinstitute.org and aregev@broadinstitute.org

**Abstract**

A vast array of transformative technologies developed over the past decade has enabled measurement and perturbation at ever increasing scale, yet our understanding of many systems remains limited by experimental capacity. Overcoming this limitation is not simply a matter of reducing costs with existing approaches; for complex biological systems it will likely never be possible to comprehensively measure and perturb every combination of variables of interest. There is, however, a growing body of work – much of it foundational and precedent setting – that extracts a surprising amount of information from highly under sampled data. For a wide array of biological questions, especially the study of genetic interactions, approaches like these will be crucial to obtain a comprehensive understanding. Yet, there is no coherent framework that unifies these methods, provides a rigorous mathematical foundation to understand their limitations and capabilities, allows us to understand through a common lens their surprising successes, and suggests how we might crystalize the key concepts to transform experimental biology. Here, we review prior work on this topic – both the biology and the mathematical foundations of randomization and low dimensional inference – and propose a general framework to make data collection in a wide array of studies vastly more efficient using random experiments and composite experiments.

Molecular biology is breathtaking in its seemingly unbounded potential for intricacy and complexity. Genomes contain as many as billions of DNA bases, encoding tens of thousands of genes, expressed in diverse spatio-temporal patterns, and acting within thousands of different cell types. In principle, the space of potential patterns, from genomes, to expression programs (regulons), stable cell types, or histological patterns, while not infinite, is enormous, although only a miniscule fraction of it is realized. A major effort in molecular biology is to discover those systems, and to decipher how their organization leads to biological function: how regulatory sequences and molecular signal encode expression patterns; how expressed genes support cellular functions; how cells partner to maintain tissue homeostasis and so on.

Many of these foundational questions are addressed by conceptually similar types of experiments – perturbing and measuring a system – in order to comprehensively chart the system's components and architecture and understand its response in different environments. In each case, we seek to establish two types of relationships: (**1**) correlative relationships that teach us which variables are similarly regulated or have similar effects, and (**2**) causative relationships that link intervention in some variables to the response of others. This is true from molecular to macroscopic scales, whether we are trying to understand how cells respond to different combinations of ligands and receptors, the set of cell-cell interactions that shape tissue structure and function, how cells and tissues respond to treatment with small molecules, the impact of and interactions between the millions of genetic variants in each of our genomes on phenotypes in health and disease, or the influence of environmental exposures on individuals and population structure.

The dramatic enhancements in our ability to perform both observations and interventions over the past decades have offered the hope that it should be possible to tackle these challenges systematically. 'Omics' in particular offered a new scale for observation, with genomics, transcriptomics, proteomics, metabolomics, etc. each collecting high-dimensional data on many molecular variables simultaneously[1]. In parallel, new perturbation methods, especially genetic perturbations, now make it possible to scale up tests of causal relationships[2]. Combining these two advances allows us in principle, and increasingly in practice, to discern, simultaneously, how each of many *observed* variables relates to condition, or how each of many *perturbed* variables relates to outcome[3,4].

There remains, however, a large gap – in some cases astronomical – between the scales of truly comprehensive experiments, and those that can be achieved in practice. This is most obvious when considering interactions, defined as non-additive effects of multiple interventions. As one example, consider signaling networks, which are commonly stimulated through combinatorial interactions. A recent, large-scale study examined pairwise interactions between 20 homodimeric ligands of the Bone Morphogenetic Protein (BMP) pathway[5]. Studying heterodimeric ligands and higher order mixtures is also of great interest in this system, especially since as many as six distinct ligands are already known to operate together in certain contexts[6]. However, doing so requires exponentially more work. If we sought to know how even one specific combination of signaling receptors would respond to any combination of 6 out of 20 BMP ligands, we would need to perform 204 times more assays than needed to survey pairwise interactions ($\binom{20}{2} = 190$ while $\binom{20}{6} = 38,760$). This gap is even more pronounced when considering genetic interactions. Each human individual, for example, carries millions of common genetic variants. Genetic association studies alone are likely to be underpowered to detect even pairwise interactions[7], and impossibly underpowered in detecting higher order interactions involving more than two loci (there simply aren't enough humans on the planet). Even if we limit ourselves to mutations previously associated with a particular disease (*e.g.*, the ~400 risk loci already associated at genome-wide significance with type 2 diabetes[8]), and assay combinations of genetic perturbations at a limited order in the lab (*e.g.*, 6$^{th}$ order, as above for BMP), comprehensive testing is inconceivable ($\binom{400}{6} \approx$ 5.5 *trillion*). If higher-order interactions each have *unique* effects, the problem indeed appears intractable.

Here, we ask: how do we choose which observations and perturbations to make, and how many of each are truly necessary to address some of these fundamental questions? A naïve view may be that each query requires a separate assay: for example, to determine the outcome of perturbing a gene, we must perform this perturbation and measure its outcome. Instead, we propose that other, far more efficient strategies may be used that jointly leverage biological knowledge and foundational mathematical principles.

In particular, we suggest that frameworks that formalize how we can leverage hidden (latent) structure in biological systems to collect data more efficiently – sometimes with exponentially greater efficiency – should be able to learn the same (or more) information. We argue that it is not necessary to measure and perturb every individual biological variable to obtain an exhaustive characterization of a system; random experiments and composite experiments often provide sufficient information, and algorithms can make accurate predictions even in non-linear systems (**Box 1**).

**Box 1: Definition of key terms**

**Structure**: Covariation, sparsity, low-dimensionality, or other characteristic patterns, typically found in high-dimensional data. Often driven by underlying biological processes.

**Latent structure**: Structure that may not be apparent in the observed representation, but is revealed by appropriate analysis.

**Latent modular representation**: A way of describing high-dimensional data through different combinations of latent structural elements (or patterns). Latent patterns might be reused across contexts (or samples) in different combinations.

**Random experiment**: Measuring or perturbing random subsets of variables or random compositions of variables.

**Composite experiment**: Measuring or perturbing entire compositions of variables (*e.g.*, a linear combination of variables).

**Group testing**: A form of composite experiment in which multiple candidates are simultaneously assayed, typically for a binary result, together in a single group, or pool. Multiple groups can be assayed, and groups can either be disjoint, or partially overlapping.

As context, we first review several precedent-setting studies in biology that demonstrate different ways we can learn much from seemingly under-sampled experiments. In each of the three case studies we present, under-sampling was a necessity, due to limited technological capabilities at the time of the experiment. We then present examples of increasingly higher-throughput approaches to making comprehensive observations and perturbations, and the discovery and utility of latent patterns that followed from these. Ultimately, we argue for an alternative strategy, where these high-throughput approaches can be used to collect data more efficiently. We will review the foundations of randomization and low-dimensional inference, and build towards a synthesis, showing that latent modular representations (**Box 1**) can be efficiently learned from random, low-dimensional experiments. Finally, we describe how these mathematical ideas can be manifested to great effect in the laboratory through random composite experiments.

**Systems biology in the pre-genomics era: under-sampling by necessity**

To systematically understand complex processes in living cells and organisms, it may seem necessary to observe 'exhaustive' examples, or at least many well-chosen ones. For instance, to optimize genotype for a desirable phenotype (*e.g.*, the catalytic capability of a particular enzyme), we might think we would need to perform saturation mutagenesis, or at least choose our perturbations by intimate knowledge of structure and function. To decipher the genetic regulatory code of gene expression in a developing organism, we might think we would need to test every sequence variant in the promoter of every gene throughout developmental time and space. To understand evolutionary relationships, we might think we would need (well-chosen) phenotypic characterizations or (exhaustive) whole genomes from many species.

Three instructive case studies demonstrate, however, that it is not, in fact, always the case that we need to observe either exhaustive or well-chosen examples (**Fig. 1**). In reviewing these results, we introduce three foundational concepts for the systematic study of biology, concerning (**i**) the ability to encode the similarity between complex, high-dimensional objects in relatively low dimensions; (**ii**) the unintuitive power and efficiency of random experiments; and (**iii**) the modularity found throughout biological systems at different levels of organization.

*Part I: A low-dimensional encoding*. Phylogenetics aims to describe a set of hierarchical evolutionary relationships between living organisms. Carl Woese, who ultimately made some of the most remarkable discoveries in the field, came to study this problem in the late 1970s. Influenced by pioneering work on molecular evolution[9,10], Woese appreciated that "[a]n organism's genome seems to be the ultimate record of its evolutionary history"[11]. However, at the time, obtaining the sequence of a complete genome of an organism was not possible, and there weren't even many sequences of specific loci. Although many phenotypic characteristics – morphology, physiology, biochemical capacity – were readily observable, it proved difficult to reconcile evolutionary relatedness inferred from phenotypes with the few DNA-based phylogenies that had been constructed at that point[12].

The insight needed to solve this problem was that one doesn't need access to entire genome sequences or even large regions of homologous genomic sequence, or to know any of the phenotypic details of each species. Evolutionary relatedness could be encoded in a short stretch of nucleotides (*i.e.*, in low-dimension) (**Fig. 1, left**), for instance in a single gene, provided the gene had a slow-evolving, relatively universal region that aligns across all species with (primarily neutral) mutations accumulating at a constant rate and serving as a molecular clock[10]. Woese correctly hypothesized that ribosomal RNAs (rRNAs) have these basic properties[11]. By comparing sequences in 16S rRNA in prokaryotes (18S in eukaryotes), Woese quickly discovered 11 new microbial phyla, and, most profoundly, an entirely new domain of life: *Archaea*[11].

Woese's work formed the foundation for the next 40 years of microbial systematics and ecology. In a broader sense, the fundamental insight – that global relatedness between a large number of apparently complex entities can be encoded in a relatively low-dimensional signal – would be applied time and again in systems biology.

*Part II: Random experiments*. In the mid 1990s, Willem P.C. Stemmer and colleagues sought to determine if naturally occurring enzymes that confer antibiotic resistance (in particular, those that are found in bacteria and are active against classic $\beta$-lactams such as penicillin) could be evolved to also confer resistance to new drugs (*e.g.*, cefotaxime, which came into commercial use in the 1980s). Testing for resistance was straightforward: a library of bacteria each carrying one mutant copy of a gene were grown in a pool in the presence of drug, the drug selected for the most resistant colonies, those were recovered, and the mutations of specific selected clones were sequenced. The critical question was, how does one choose the candidate mutations to test? Testing every conceivable mutant sequence was clearly impossible (since the number of unique sequences explodes combinatorially, exceeding the number of bacteria that could be assayed), and selecting a 'rational' panel of mutations required (**i**) hard-earned knowledge of structure and function, and **(ii)** the ability to generate sequences with defined mutations, which was not readily possible at scale at that time.

Stemmer instead developed DNA shuffling (**Fig. 1, middle**), a method to easily generate large libraries consisting of random mutations in the starting gene. DNA shuffling[13,14] would ultimately play a key part in a coming revolution in bioengineering and the development of directed evolution[15]. In its original formulation, DNA shuffling was used to generate a library of tens of thousands of mutated copies of a gene, by random fragmentation and PCR-based recombination[13]. Stemmer and colleagues went on to develop a more powerful DNA shuffling approach, in which the initial pool of genes included orthologous copies from multiple species[14], and candidates combined point mutations and random chimeric fragments from the different species. Iterative rounds of mutation and selection were then used to evolve the gene *in vitro*. After just a few rounds of selection, clones evolved that were thousands of times more resistant than the original. Interestingly, different selected clones had shared subsets of mutations, while also each containing an excess of mutations that were shown to be unnecessary by backcrossing with wild-type genes.

How do we explain the remarkable success of these random experiments, especially as no knowledge of the structure or function of the enzyme was needed in order to improve that enzyme? First, opting for random mutants made it very easy to generate a large candidate library covering highly diverse sequences, and the pooled screening format meant that each of these could be tested in a single assay. Second, and especially with multi-gene shuffling, the candidates consisted of mutations in an existing protein or new combinations of functional gene fragments, such that more of the library likely produced functional protein (compared with completely arbitrary sequence).

In addition, each candidate consisted of many more mutations than likely contributed to the optimized functionality. This, somewhat unintuitively, increased the efficiency of the method. For instance, in the first screen, the best clone had 9 mutations, but at most 5 of these were necessary (about 1 in 3 random mutations are expected to be silent). By generating mutations at a higher order per candidate (*e.g.*, 9), each candidate in effect simultaneously tested many lower order combinations. Thus, in addition to being efficient in terms of execution and producing functional proteins, these random experiments achieved efficiency by simultaneous group testing (**Box 1**) of many perturbations.

*Part III: Inherent modularity*. Around the same time, developmental biologists were pursuing foundational systematic studies of the *cis*-regulatory circuitry of gene expression. Their goal was to determine how DNA sequences, and the transcription factor binding sites they encode, direct organismal development through control of gene expression. As a primary tool, they used gene transfer experiments, in which regulatory sequences cloned together with a reporter gene were injected into a fertilized egg or developing embryo, followed by an *in situ* assay along development to determine the spatial expression patterns driven by the regulatory sequences. The most extensive such experiments were conducted at the time in sea urchin[16,17], with similar early efforts in fly and mouse[18–20] (as reviewed in[21]).

In principle, perturbation to each DNA sequence element could have had a unique effect in every genomic, temporal, and spatial context. However, it became clear early on that the developmental systems studied were characterized by multiple layers of latent modularity (**Fig. 1, right**). At the sequence level, the regulatory regions upstream of developmentally important genes were comprised of modular elements, each consisting of specific, separable sequences of DNA with distinct regulatory functions[22]. At the gene level, there were co-regulated modules, or regulons, with shared modular sequences in their *cis*-regulatory regions. Finally, at the level of the embryo, there were spatial modules of (co-) expression patterns that could be readily connected to canonical structures of the developing organism. In a dramatic demonstration of modularity, two sequence elements driving distinct spatial expression patterns could be combined together in a single regulatory region, resulting in a pattern identical to the sum of the individual patterns[23,24].

Ultimately, the largest scale coordinated efforts in these early studies culminated in the description of a regulatory network of 40 genes directing the development of the sea urchin embryo[17]. Soon, and as envisioned at the time[25], genome scale technologies made systematic studies and the discovery of latent patterns possible at far greater scales. This would happen with both the completion of whole genome sequencing efforts, which, among other things, provided a "parts list" of the fundamental constituent elements in any system, as well as with the development of multiple modes of perturbation and genome-scale measurements (such as transcriptomics and proteomics), which would facilitate the large-scale elucidation of gene modules, and enable studies of *trans* regulation.

**Latent modular patterns recovered from comprehensive genomic observations**

The catalog of genes identified for each species from whole genome sequences[26–28] was not only a critical stepping stone to systematic measurements, but the vast amounts of generated data allowed the new field of computational genomics to search for and find recurrent latent sequence patterns: unknown upfront, but recovered by algorithms from the data. Their discovery relied on commonalities but also allowed to assess unexpected deviations at multiple levels, either between individuals of the same species, or across the "reference" genomes for multiple species. For example, while protein coding genes could be identified partly based on known patterns derived from existing biological knowledge (*e.g.*, start and stop codons), *cis*-regulatory motifs were typically discovered *de novo*, by finding latent patterns through their recurrence across many examples. This was facilitated by grouping orthologous regions in different species[29] or across regulatory regions adjacent to genes known or expected to be co-regulated, co-expressed or functionally related[30–32].

Co-*regulated* modules of genes whose expression levels co-vary across samples or conditions, and often are controlled by shared *cis*-regulatory motifs and TFs, were one of the most predominant patterns that emerged from the growing catalogs of genome-scale expression profiles (**Fig. 2**). Even in the earliest results, co-regulated gene modules were observed in many contexts: expression profiles in cycling cells revealed oscillating modules in the different phases of the cell cycle[33]; modules associated with different metabolic processes in cells grown in different conditions[34]; modules in various (sub)types of cancer that reflected differences in the functional demands of each tissue and dysregulation in tumors[35–37]; and developmental programs revealed as transcriptional 'waves' during differentiation[38]. These co-regulatory modules could be organized into loose hierarchies, reflecting more global and specialized cellular processes, and were congruent with modules defined from latent patterns on other biological data, such as physical interaction between proteins or co-occurrence across genomes in evolution[39,40]. The core principles of these analyses – in particular, the kinds of inferences we can make about biological processes from coordinated expression across contexts – remain relevant today, with single cell genomics leading to the discovery and characterization of ever finer cellular programs[41].

**Systems biology through perturbation at scale**

Observational approaches to infer gene regulation were complemented by perturbational approaches, targeting individual genes and gene pairs, which directly enable causal inference and discovery of some aspect of the co-regulatory mechanism. The most extensive genetic perturbations studies to date have been conducted in yeast, due to the facility of genetic manipulation, crossing, and pooled screens[42–44]. Just as **measuring** many genes across growth conditions revealed **co-regulated** modules, testing the impact of **perturbation** (deletion in this case) of many genes across multiple conditions revealed **co-functional** modules that were, for example, required for galactose utilization[45], or regulation of the response to high osmolarity, among others[44] (**Fig. 2**). Over time, technological developments, in particular RNAi[46] and CRISPR/Cas9 genome editing[47–49], enabled genome-wide screens for essentiality[50,51] or marker gene expression[52] in multiple organisms, including mammals. These approaches have since been adapted to study genes critical to diverse biological processes; for example, those dispensable for cell growth, but required for HIV cell entry[53].

Analysis of genetic interactions – the extent to which joint perturbation of multiple genes (typically two) causes a different phenotype than that expected by the additive impact of their individual perturbations – allowed discovering co-functional modules of genes in new ways, based on the similarity in their genetic interaction profiles (*i.e.*, the vector of interactions of one gene with all other genes). A series of studies[54,55,56] queried interactions at a near genome-wide scale in yeast, with all possible pairwise deletions of ~6,000 genes. From the ~23 million pairs tested to date, around 1 million significant interactions were found. Hence, 2-way gene interactions are both relatively sparse (~4.3%), and highly prevalent in absolute terms. The same pattern was found for 3-way gene interactions[57], which are estimated to be ~100-fold more prevalent in absolute terms while being ~3-fold sparser in relative terms. Genes that were previously known to function in the same pathways had similar interaction profiles, forming co-functional modules, with the most similar profiles clustered into specific complexes and pathways, which further clustered into distinct biological processes (such as DNA repair), and finally into different cellular compartments[57]. This organization remarkably mirrors that obtained from module-level analysis of gene co-expression patterns, suggesting that co-functional and co-regulated modules were often coherent. Notably, many genes had pleiotropic functions – they interacted with a functionally diverse set of genes – and, based on co-functional modularity, it was possible to infer new functions for genes with no prior annotations[56].

Most of these genetic screens have relied on simple readouts, such as essentiality or a gene marker, but because these are one-dimensional and integrate effects from multiple processes, it remains to be determined exactly what each perturbation actually does to the cell. To understand that, without *a priori* knowledge, requires a high-content measurement. Studies in yeast did so by combining systematic perturbations with global measurements of gene expression profiles. These experiments perturbed components of either a specific known system (such as the galactose utilization[45] or the response to osmotic shock[58]) or spanning a broad range of yeast genes (as in the Rosetta Compendium[59]), and found that measured genes can be grouped into co-regulated modules, and perturbed genes could be similarly grouped to reveal co-functional modules. With advances in RNAi (and later, CRISPR/Cas9) based perturbation, combining perturbation with profiling was extended to mammalian cells, for example with readout based on high-content imaging[60], or using expression to dissect transcriptional networks that mediate innate immune response to a diverse set of pathogenic stimuli[52,61] or differentiation of adaptive immune cells[62].

In contrast to genome-wide screens of viability or an expression marker, studies with a transcriptional or imaging readout were historically limited by scale – because each perturbation had to be assayed separately. With the advent of single cell genomics, methods like Perturb-Seq recently paired pooled CRISPR perturbations with single-cell RNA-seq to substantially increase the throughput of such experiments, as well as to enable characterization of heterogenous response to individual perturbations[4,63,64]. The subset of combinatorial perturbations that have been analyzed in this way highlight the value of a high-content readout by showing that the same combinatorial perturbation can have different effects on different genes[4,65]. More recently, CRISPR perturbations have been paired with a high-content optical readout[66,67], enabling even larger scale studies[67].

Thus, from both observational and interventional studies, there is extensive modularity both in terms of co-regulated and co-functional genes. The most recent technological advances have begun to chart the way to have both larger numbers of perturbations and a high dimensional measurement; these should be especially valuable for understanding the impact of combinations of perturbations.

**The mathematics of learning high-dimensional parameters from low-dimensional data**

Technology will no doubt continue to progress, and there is much we will learn by continuing with the current paradigm (including, for example, refining our notion of what exactly a gene module is). However, as we noted at the outset, the number of hypothetical possibilities, especially for genetic combinations, remains astronomical. Thus, it is worth asking: given that the number of modules we tend to discover is so much smaller than the global scale of measured or perturbed variables, could we somehow collect less data, and yet learn the same (or more) information? To this end, we review the mathematical foundations of this problem, developed independently of biology, towards a synthesis of how modular patterns can be efficiently learned from random, low-dimensional experiments.

Recall the three seminal experiments that introduced three foundational concepts for systematic studies in biology (**Fig. 1**): (**i**) the ability to encode the similarity between complex, high-dimensional objects in relatively low dimensions (Carl Woese and the study of orthologous genes to build the tree of life); (**ii**) the power and efficiency of random experiments (DNA shuffling and random mutagenesis in directed evolution); and (**iii**) the modularity that seems inherent at multiple biological scales (modular sequences, genetic controls, expression programs, and spatial patterning in the embryo). In fact, such experiments are successful because of specific mathematical principles.

The mathematical foundations of modern experimental design trace back to George Box and R.A. Fisher[68,69]. Fisher, in particular, evangelized for randomization in experimental design as a way to mitigate systematic error. Consider, for example, an experiment designed to test for differences in crop yields between two treatments. Suppose we place all of Treatment 1 on plots at the east end of a field, and all of Treatment 2 on plots at the west end (or vice versa). If there is a confounding correlation between east-west placement and crop yield, then the inferred difference between treatments will either be far too low (**Fig. 3a**), or far too high (**Fig. 3b**). Thus, one should randomize the placements of each treatment across the field.

In our case, we are interested not just in systematic bias, but also in efficiency. Abstractly, we can think about efficiency in terms of dimensionality: computing in low dimension is more efficient than in high dimension, just as collecting low-dimensional data may be less work, faster, or less costly than collecting high-dimensional data. Thus, we ask a more generalized question: suppose we have data that could be drawn from a mixture of two distributions (as in **Fig. 3c**); in general, can we embed these data in low-dimension, while maintaining separation between the distributions?

The answer to this question derives from a seminal result of Johnson and Lindenstrauss[70] that states that any set of $n$ points in Euclidean space can be embedded into $O(\frac{\log(n)}{\epsilon^2})$ dimensions, while preserving all pairwise distances up to a factor of $1 \pm \epsilon$. (We will discuss a specific example below, with more biological applications in the next section.) In particular, this result does not depend on the original dimensionality of the points, because we only need enough dimensions to encode the similarity of the points, and nothing about what those points actually represent. This property underlies the success of Woese's approach: a single set of orthologous genes has enough nucleotides to encode a large number of phylogenetic distances.

Clearly, we cannot hope for distances to be preserved in any embedding. So how do we choose a good embedding? Interestingly, standard data-driven dimensionality reduction techniques, such as principal components analysis (PCA), are not guaranteed to work. For instance, if we project data drawn from the distributions in **Fig. 3c** onto the first PC, the two distributions will collapse into one. Conversely, and somewhat surprisingly, random linear projections preserve the geometry of points with high probability[71,72].

There are many examples and extensions of the idea that random projections can provide a good embedding. Most relevant to this example, data drawn from $k$ high-dimensional Gaussians can be randomly embedded into $O(\log(k))$ dimensions, while maintaining separation between the clusters. On the other hand, if the Gaussians are highly eccentric, like our 2-dimensional example above, PCA with $\log(k)$ components might collapse some of these distributions together[73]. Random linear projections can even preserve nonlinear geometry. For instance, distances (and geodesic distances) on a $k$-dimensional, well-conditioned submanifold in $n$ dimensions are preserved by approximately $O(k \log(n))$ random projections[74].

We now connect this back to randomization in experimental design and R.A. Fisher's crops. Suppose that the high-dimensional data consist of the output of each individual plant, each of which has received one of $k$ treatments. The goal is to determine if there is a significant difference in output between the treatments. To gather the high dimensional data, Fisher (or his associates) would have had to march through the field and measure the output of every plant, one-by-one, and then conduct all calculations by hand. Hence, they were quite motivated to work in low dimensions, by looking at the net (or average) output of entire plots, each containing, say, hundreds of plants with a single treatment. We can think of these measurements as linear projections of the high-dimensional data, $y = Ax$, with $x \in \mathbb{R}^p$ the output of every plant, $y \in \mathbb{R}^l$ the output of every plot, and $A \in \mathbb{R}^{l \times p}$ defining which plants are in which plots. In a typical layout, plants in a plot would be contiguous, and $A$ would have a blocky design. Moreover, in the original (pathologically bad) configuration, the plots for each treatment were also contiguous, and, thus, could either collapse or amplify differences (as in **Fig. 3a,b**). Conversely, if each plot were randomly defined, there is a very low probability of realizing the pathologically bad configuration, and, with high probability, $O(\log(k))$ plots would suffice to maintain separation between $k$ treatments. (This raises issues of practicality for the farmer, but will suffice to illustrate the point.)

Going further, it is possible to build efficient machine learning algorithms, for example, for classification or regression, using randomly-defined features. That is, in many cases we can build models that use as input a small number of randomly defined features rather than native high-dimensional, engineered, or learned features, and expect similar performance. Random features can be used in kernel machines[75], or in a more generalized way in random 'kitchen sinks'[76], and convolutional networks with random filters can perform surprisingly well[77].

The most celebrated example of learning from random projections is compressed sensing[78,79] that recovers high-dimensional values from a set of low-dimensional observations, *i.e.*, in $y = Ax$, given observed low-dimensional data $y$ (*e.g.*, several composite measures of abundance, intensity, etc.) and compositions $A$, solve for the unobserved high-dimensional values of interest $x$. Without constraints, it is impossible to identify a unique solution to this problem, since $A$ is not invertible. The insight of compressed sensing is that the problem becomes tractable if $x$ has few degrees of freedom (*e.g.*, $x$ is $k$-sparse, with only $k$ nonzero values), and $A$ satisfies certain conditions, formalized by the restricted isometry property (RIP)[80]. Critically for our applications, $x$ can be naively dense, so long as it has a sparse representation in some basis. For instance, if $x$ can be represented by a sparse combination of $k$ features in a dictionary (*i.e.*, $x = Dw$ with $D \in \mathbb{R}^{m \times n}$ and $w$ being $k$-sparse), then only $O(k \log(n))$ random projections are needed to recover $w$ and, hence, $x$ (assuming $AD$ has nice RIP constants).

The central idea of compressed sensing – using the knowledge that a certain category of structure exists to recover high-dimensional parameters from low-dimensional observations – has been extended to several related problems. In matrix completion, for example, one observes a small fraction of randomly-selected entries from a matrix, and then tries to recover the unobserved values. With prior knowledge that the matrix is rank $r$, it can be shown that only $O(n^{1.2} r \log(n))$ observations are needed to recover all $n \times n$ values[81]. Similar results apply to recovering higher-order tensors[82]. Interestingly, compressed sensing and related under-sampling problems exhibit phase transitions in successful recovery of high-dimensional parameters: recovery is highly likely to succeed anywhere above a boundary on the number of observations, while being highly likely to fail anywhere below the boundary[83].

In summary, we draw a thread from Fisher to Johnson and Lindenstrauss to help us think about randomization and low-dimensional experiments. Then, compressed sensing teaches us that we can recover the latent patterns of a high-dimensional signal from random, low-dimensional data. We can now turn back to how we can leverage what we can assume about structure in biology to design new experiments in lower dimension and recover more information to solve challenging biological problems.

**A framework for random compressed experiments in biology**

These mathematical concepts offer new ways to tackle studies of genetics, molecular biology and gene regulation at the necessary scale in an experimentally tractable manner. Moreover, the variety of studies we discussed above illustrates the broader diversity of biological problems that may also be tackled using these methods.

We reviewed above biological and mathematical precedents for three central aspects of our framework (**Fig. 4**):

(1) *Learning in low dimension*. Low-dimensional embeddings can efficiently encode the geometry of high-dimensional data. This means, for example, that the number of clusters of co-expressed genes or samples we can resolve scales almost exponentially with the embedding dimensionality. Thus, for certain learning tasks that don't explicitly depend on high-dimensional details (*e.g.*, clustering, estimating similarity, or classification based on expression profiles), we only need to collect low-dimensional data, as opposed to collecting high-dimensional data, then projecting into low dimension for computational efficiency or robustness.

(2) *Random embeddings*. While it is possible in some cases to construct efficient data-dependent embeddings, random low-dimensional embeddings preserve geometry in a general fashion.

Moreover, for certain experiments, random embeddings can be far easier to implement in the lab than pre-determined designs.

(3) *Recovery of modular patterns*. For high-dimensional data that are known to have a sparse, modular representation, such as co-regulatory modules or co-functional modules, we can recover the modular activity (and estimate the high-dimensional signal) from random, low-dimensional data. In some cases, prior knowledge of sparse modularity is sufficient for this recovery, even in the absence of any training data.

Implementing these ideas is especially compelling in the context of learning regulatory networks, where we wish to find not only how genes respond but also which other genes control their responses, and where combinatorial explosion in the number of possible genetic interactions between regulators makes it impossible to conduct exhaustive experiments. We thus describe how we can solve this problem by applying the ideas above to efficiently study co-regulated and co-functional modules of genes. To this end, we will assume that there are $g$ gene expression levels to measure, $p$ perturbations to make (targeting either individual targets or multiple targets in combination; we discuss nonlinear interactions in more detail below), $c_r$ co-regulated modules, and $c_f$ co-functional modules.

Experimentally, there are two ways to collect low-dimensional data: by sampling, or by composition (**Fig. 4b**). With sampling, subsets of elements are profiled (*e.g.*, measuring $g$ expression levels for only a sampled subset of $p$ perturbations). With compositions, multiple elements are profiled simultaneously (*e.g.*, by measuring linear combinations of the $g$ genes in each of $p$ perturbations). A composite measurement reflects a weighted abundance, and a composite perturbation produces a weighted outcome. Samples can be chosen at random, and compositions can be randomly-defined. For compositions, there are multiple random distributions that can be used in low-dimensional embeddings, including Gaussian, Bernoulli, Rademacher, and sparse variations thereof[71,73,84].

The key power of pursuing random compositions or random samples becomes apparent when we consider how the number of measurements or perturbations approximately needs to grow relative to the size of the inference task (**Box 2**): the necessary number of experiments often scales logarithmically in the total number of latent patterns (*e.g.*, as $\log c_r$ or $\log c_f$), and linearly in the size of the modular representation (*i.e.*, the number of co-regulated or co-functional modules needed to describe a single high-dimensional signal) (**Fig. 5**).

**Box 2: Approximate scaling of random compositions or random samples for several common tasks**

> We first note that we can learn about the modular architecture of a biological process from data collected in an *unknown* random low-dimensional embedding. For instance, if we measure $g$ genes in $O(\log c_r)$ random perturbations of unknown composition, we can still correctly infer co-regulated clusters. This formalizes the notion that we can extract meaningful clusters of genes from a panel of samples without knowing what was specifically different about each sample. Similarly, we can measure $O(\log c_f)$ random and unknown compositions of genes in $p$ perturbations, and still infer clusters of co-functional genes. That is, we can make random measurements of unknown composition of the outcome of perturbations, yet still accurately infer perturbation clusters. This would be similar to observing a (possibly high-dimensional) cell biology phenotype whose molecular nature is unknown, and characterizing the perturbed genes by their impact on this phenotype[85].

However, for biological understanding, we ultimately do want to decompress the low-dimensional data, recover modular activities, and estimate the unobserved high-dimensional values. Suppose first that we assume the response of each gene to a panel of each of the $p$ perturbations can be described by a $k_f$-sparse combination of responses to the $c_f$ co-functional modules – in other words, we assume that even if any given gene is affected by many perturbations, it is ultimately regulated by relatively few co-functional modules. We can also assume we have previously learned these modules, for example using the approach with random projections outlined in the previous paragraph; the goal now is to learn which co-functional modules regulate each gene.

If the modular assumptions are true, then we can approximately recover the response of each gene to every perturbation from $O(k_f \log c_f)$ random composite perturbations, the outcome of each corresponding to a random linear combination of outcomes to individual perturbations. While we discuss experimental implementations in the next section, in essence, we would measure the sum (bulk) outcome of multiple perturbation experiments.

Similarly, suppose that the outcome of each perturbation in terms of the response of $g$ genes can be described by a $k_r$-sparse combination of the $c_r$ co-regulated modules. If this is true, then we can approximately recover the outcome of each perturbation from $O(k_r \log c_r)$ random composite measurements (as discussed below, here we will measure the net abundance on multiple genes).

In either case, the composite approach is nearly exponentially more efficient than performing each perturbation or measurement individually (**Fig. 5**).

In other cases, we might prefer to collect low-dimensional data by random sampling. This might be most natural when studying combinatorial perturbations. For instance, if the outcomes to all $O(p^3)$ three-way perturbations of $p$ targets is rank $c_f$, then we can estimate all effects by measuring the outcome of $O(p^{1.5}c_f)$ randomly-sampled three-way perturbations[86]. Compared with the scaling above ($O(k_f \log c_f)$ composite perturbations), random sampling might therefore be far less efficient, but might nonetheless be more appropriate, or at the very least complimentary, in certain contexts. We address these in the next section, in which we discuss experimental implementations of random compressed experiments.

**Experimental implementations of compressed experiments**

The three seminal experiments that we reviewed from the pre-genomics era established experimental precedents for under-sampling that have since been advanced in many ways. To highlight a few examples: compressed microarrays have been used to test for differential expression[87]; random DNA probes have been used for microbial diagnostics[88]; composite metagenomic sequencing can be decompressed into individual genomes[89,90]; *de novo* energy potentials have been derived using methods of compressed sensing to predict protein-DNA specificity[91]; pooled sequencing designs were used to reduce genotyping costs[92]; random sequences have been used to train *cis*-regulatory models at massive scale[93,94]; single-cell regulatory states can be deconvolved from bulk RNA samples[95]; shallow single cell RNA-Seq is used to make inferences about cell type, cell state and cellular programs[96–99]; fluorescent images can be super-resolved to nanometer scale using randomness and sparsity in fluorophore emitters[100]; and computation can be scaled by computing directly on compressed data[101]. We posit that it is now possible to progress to new compressed experimental designs, for both measurement and perturbation, that follow directly from the framework we presented.

There are several ways to implement composite measurements of gene expression. We have previously showed[102] that qPCR can be used to generate composite data by measuring the net signal produced from simultaneous amplification of multiple genes. Although this method is not likely to be easily scaled, it served as a proof of concept. Sequencing could be used to generate composite data in higher throughput by attaching barcodes during PCR to cDNA from different genes included in a composition. The 'abundance' of each composite measurement could be determined by counting reads for each composite barcode, although, because sequencing is a digital assay, the primary benefit would be to enable sequencing of shorter reads (rather than needing reads long enough to uniquely map to each gene, reads would only need to be long enough to map each barcode). As the primary effect is to make multiplexing more efficient, composite measurements might have the most value in multiplex-limited settings, like flow cytometry or fluorescent *in situ* hybridization (FISH). In these cases, genes in a composition could first be targeted by oligonucleotide probes containing target hybridization sequences in addition to a barcode, followed by hybridization of fluorescent probes to the barcode, with the net fluorescent signal for a barcode corresponding to the composite measurement. We have recently implemented such a method, showing that composite measures of RNA abundance can make methods of imaging transcriptomics at least two orders of magnitude more efficient[103].

As multiplexing and tissue volumes increase, ultimately aiming for entire organs, we expect gains in efficiency to improve by several additional orders of magnitude, facilitated by the same mathematical principles and composite measurement framework applied to different experimental aspects. For example, imaging ~140 genes (the size of a standard MERFISH[104] experiment) throughout an entire mouse brain (size 400 mm$^3$) with standard multiplexed FISH at 60X magnification, would require ~67 million fields of view, corresponding to ~1 billion images across 16 rounds of imaging, generating 12 petabytes of data and taking ~10 years of imaging time on one instrument. (For those not yet dismayed, note that a human brain is ~1,000 times larger.) In contrast, with a composite approach we can tackle multiple points of inefficiency: reducing the rounds of imaging by probing for linear combinations of gene abundance (as described above); reducing scanning time by imaging at low resolution and applying principles of super resolution (with each low resolution pixel representing a composition of high resolution pixels defined by a point spread function); and, reducing the number of colors that must be separately imaged by exciting the sample with a composition of wavelengths.

Composite measurements can also be made using alternative high-throughput technologies. Of immediate interest will be composite *in situ* proteomics, for which there are multiple, recently-developed *in situ* profiling methods, including MIBI[105], CODEX[106], and Immuno-SABER[107]. We can also imagine composite ChIP-Seq or composite ATAC-Seq to study chromatin, composite measures of metabolites for cheap, yet informative, diagnostics, or composite methods of live-cell imaging, using, for example, Raman scattering[108].

Unanticipated nonlinear effects within composite measurements are a primary concern with any such method. *In situ* proteomics, for example, spans a very large potential range of abundances, and imaging at an intensity that is appropriate for low-abundance signal might result in saturated (nonlinear) high-abundance signal. There are both molecular and computational ways to mitigate such problems. We can molecularly normalize anticipated highly-abundant signals, using 'cold' competing probes in proportion to prior estimates of abundance, or normalize with differential signal amplification (*e.g.*, using SABER[109] or Immuno-SABER[107]). Computationally, methods of 1-bit compressed sensing[110,111] can robustly recover sparse modular patterns (up to a scaling factor) from highly-quantized, even binary (presence / absence), data.

Like composite measurements, there are multiple conceivable ways to implement composite perturbations. Many of these are positioned to leverage 'natural' manifestations of random distributions in high throughput experiments. It is possible, for example, to randomly introduce many perturbations per cell – for instance, by infecting with an sgRNA library at high MOI[112] such that each cell receives a random combination, or using vectors with multiple perturbations per construct[113] – while aiming to learn lower-order nonlinear effects; each cell would thus represent a group test of many lower-order perturbations. (With CRISPR perturbations this might necessitate inhibition rather than knock-out, as cells will only tolerate so many double-stranded breaks.) We can also randomly 'overload' cells, each with different perturbations but the same barcode, in droplet[114] or split-pool[115] library preparations.

Beyond composite genetic perturbations, composite experiments could be similarly applied across a wide array of biological problems for which we currently lack a fundamental, comprehensive understanding of the nature of interactions. For instance, in microbial ecology we might wish to test combinations of bacterial strains for total biomass production, selection for a particular strain, or net production of a metabolite. It will be impossible to test every complex (higher order) community of interest, but we can use composite experiments to first construct random higher order combinations of strains – for example, by merging random subsets of barcoded droplets, each containing a single strain[116] – then measure the outcome of each composition, and finally infer the outcomes of experiments with less complex communities. Or, we could perform composite drug screens to identify interesting candidates for combination therapy. This might be accomplished through barcoded droplets, or perhaps barcoded drug-containing beads, in an *in vitro* setting, though controlling (or measuring) dose may pose a challenge. Relatedly, lower-order combinations of extracellular ligands that lead to desired outcomes (e.g., in cellular differentiation or organoid development) could be identified through higher-order composite experiments.

With higher-order perturbations and the goal of learning lower-order outcomes, unanticipated nonlinear effects are a concern. Specifically, if we perturb at order $n$ while aiming to learn nested effects at order $m < n$, any nonlinear effects of perturbations at orders higher than $m$ will potentially confound our results. Studies of genetic perturbations in yeast suggest that higher-order effects are increasingly rare (though prevalent in absolute terms)[57], though the true density for most biological systems remains to be determined.

We can additionally take a complimentary approach using random sampling. Random samples of combinatorial perturbations can be used to infer the effects of unobserved combinatorial perturbations of the same order. For instance, using the outcomes to a random sample of pairwise genetic knockouts, matrix completion can be used to infer the effect of every pairwise knockout in a panel of targeted genes[65]. The same idea applies more broadly to "combinatorial conditions" where each member of a given class is assayed in combination with members of another class, and the full matrix of outcomes is assumed to be low-rank. If we wish to determine the neutralizing capacity of a panel of antibodies against a panel of pathogens, for example, we could randomly sample pairs of antibodies and pathogens, and use these assay results to infer the full matrix of effects. In each case – for combinatorial perturbations and conditions – the ideas can be generalized to higher-order (*i.e.*, multiple perturbations or multiple classes) using tensor completion. Finally, the success of random, sparse sampling in "under-sequenced" single-cell RNA-Seq in inferring salient low-rank structure[98] can be extended to emerging single-cell epigenetic profiling techniques with inherent limitations on low copy-number per cell.

**Random compositions as natural elements of biological systems**

Just as experimentalists can use the principles of random compositions to efficiently study a system, natural systems might employ random compositions to efficiently perform a given function. For instance, many neural systems are characterized by converging projections through successive hierarchies (*e.g.*, the projections into higher brain regions from glomeruli representing different odorant receptors[117]). It is possible that these converging projections are exquisitely arranged to represent specific combinations, but as we discussed above this is not necessary, at least for certain computational tasks. In order to form a faithful neural representation of similarity and dissimilarity of environments at the point of converging neurons, random projections from lower brain regions may suffice (just as the random composite measurements above are sufficient to cluster samples by similarity). More generally, any biological computation that relies only on the geometry of inputs might utilize a physical manifestation of random low-dimensional embeddings.

These principles might also appear in more 'passive' roles. In population genetics, for example, many variants causally associated with complex, polygenic disease reside in regulatory regions, suggesting that they act through regulatory networks[118,119], and recent studies reveal modular patterns in GWAS signals that are linked with cellular physiology[120,121]. Our framework for random composite experiments can inform the study of these variants. In particular, we can consider each genome as a random composite perturbation (each individual has millions of common variants[122]), and attempt to infer how modules of genetic variants affect phenotype. This could be done either allowing for nonlinear interactions, or accounting only for linear effects. Since we might be underpowered to learn the modular organization of variants from human genetics alone, especially for nonlinear interactions, we can use prior knowledge of co-functional modularity gained from cellular studies to complement human genetic approaches. In doing so, we might gain greater mechanistic insight into the genetics of polygenic disease, and we might make GWAS more statistically powered by testing for association at the level of dozens or hundreds of modules, rather than millions of individual variants.

A natural extension to the random composite, modular view of human genetics, is to view evolution as a dynamic series of random composite experiments. In each generation, each individual realizes a random composition of mutations. These individuals are then subjected to a composition of selective forces, just as the environment is modified by a composition of species. While evolutionary dynamics certainly differ from the non-adaptive experiments above, exploring the principles of low-dimensionality, randomness, and modularity may nonetheless lead to new insights about rates and points of convergence, and the unexpected efficiency of a randomly-evolving system.

**Outlook**

Over the past several decades there has been increasing use of algorithms as *applied tools* in the biological sciences. The success of these approaches has paved the way for a new paradigm, in which we view biology through an algorithmic *lens*. Using an algorithmic lens we can ask, what are the limits of biological computation, and what are the limits of what we can learn about biological processes? Answers to these questions can inspire new modes of experimentation and discovery, just as the new data and hypotheses we generate can inspire new directions in algorithmic research. Our hope is that the framework presented above becomes part of a larger, global paradigm shift towards the algorithmic lens on the study of life.

**Acknowledgements**: We thank A. Hupalowska for help with figures, and K. Shekhar and P. Blainey for helpful discussion and feedback. Work was supported by the Klarman Cell Observatory and HHMI (AR), and by the Merkin Institute Fellowship (BC). **Declaration of interests**: Aviv Regev is a co-founder and equity holder of Celsius Therapeutics, an equity holder in Immunitas, and was an SAB member of ThermoFisher Scientific, Syros Pharmaceuticals, Neogene Therapeutics and Asimov until July 31, 2020. From August 1, 2020, A.R. is an employee of Genentech. B.C and A.R. are named as inventors on a patent application related to composite measurements.

**References**

1. Joyce, A. R. & Palsson, B. The model organism as a system: Integrating 'omics' data sets. *Nature Reviews Molecular Cell Biology* (2006). doi:10.1038/nrm1857

2. Shalem, O. *et al.* Genome-scale CRISPR-Cas9 knockout screening in human cells. *Science* **343**, 84–7 (2014).

3. Liberali, P., Snijder, B. & Pelkmans, L. Single-cell and multivariate approaches in genetic perturbation screens. *Nature Reviews Genetics* (2015). doi:10.1038/nrg3768

4. Dixit, A. *et al.* Perturb-Seq: Dissecting Molecular Circuits with Scalable Single-Cell RNA Profiling of Pooled Genetic Screens. *Cell* **167**, 1853-1866.e17 (2016).

5. Antebi, Y. E. *et al.* Combinatorial Signal Perception in the BMP Pathway. *Cell* (2017). doi:10.1016/j.cell.2017.08.015

6. Simic, P. & Vukicevic, S. Bone morphogenetic proteins in development and homeostasis of kidney. *Cytokine Growth Factor Rev.* (2005). doi:10.1016/j.cytogfr.2005.02.010

7. Zuk, O., Hechter, E., Sunyaev, S. R. & Lander, E. S. The mystery of missing heritability: Genetic interactions create phantom heritability. *Proc. Natl. Acad. Sci. U. S. A.* (2012). doi:10.1073/pnas.1119675109

8. Mahajan, A. *et al.* Fine-mapping type 2 diabetes loci to single-variant resolution using high-density imputation and islet-specific epigenome maps. *Nat. Genet.* (2018). doi:10.1038/s41588-018-0241-6

9. Zuckerkandl, E. & Pauling, L. Molecules as documents of evolutionary history. *J. Theor. Biol.* (1965). doi:10.1016/0022-5193(65)90083-4

10. Kimura, M. Evolutionary rate at the molecular level. *Nature* (1968). doi:10.1038/217624a0

11. R. Woese, C. R. & E. Fox, G. E. Phylogenetic structure of the prokaryotic domain: the primary kingdoms. *Proc. Natl. Acad. Sci. U. S. A.* (1977).

12. Knight, R. *et al.* The Microbiome and Human Biology. *Annu. Rev. Genomics Hum. Genet.* (2017). doi:10.1146/annurev-genom-083115-022438

13. Stemmer, W. P. C. Rapid evolution of a protein in vitro by DNA shuffling. *Nature* (1994). doi:10.1038/370389a0

14. Crameri, A., Raillard, S. A., Bermudez, E. & Stemmer, W. P. C. DNA shuffling of a family of genes from diverse species accelerates directed evolution. *Nature* (1998). doi:10.1038/34663

15. Arnold, F. H. Combinatorial and computational challenges for biocatalyst design. *Nature* (2001). doi:10.1038/35051731

16. Kirchhamer, C. V & Davidson, E. H. Spatial and temporal information processing in the sea urchin embryo: modular and intramodular organization of the CyIIIa gene cis-regulatory system. *Development* (1996).

17. Davidson, E. H. *et al.* A genomic regulatory network for development. *Science* **295**, 1669–78 (2002).

18. Hoch, M., Schröder, C., Seifert, E. & Jäckle, H. cis-acting control elements for Krüppel expression in the Drosophila embryo. *EMBO J.* (1990).

19. Gomez-Skarmeta, J. L. *et al.* Cis-regulation of achaete and scute: Shared enhancer-like elements drive their coexpression in proneural clusters of the imaginal discs. *Genes Dev.* (1995). doi:10.1101/gad.9.15.1869

20. Frasch, M., Chen, X. & Lufkin, T. Evolutionary-conserved enhancers direct region-specific expression of the murine Hoxa-1 and Hoxa-2 loci in both mice and Drosophila. *Development* (1995).

21. Arnone, M. I. & Davidson, E. H. The hardwiring of development: organization and function of genomic regulatory systems. *Development* (1997).

22. Kirchhamer, C. V., Yuh, C. H. & Davidson, E. H. Modular cis-regulatory organization of developmentally expressed genes: two genes transcribed territorially in the sea urchin embryo, and additional examples. *Proc. Natl. Acad. Sci.* (2002). doi:10.1073/pnas.93.18.9322

23. Gray, I., Szymanski, P. & Levine, M. Short-range repression permits multiple enhancers to function autonomously within a complex promoter. *Genes Dev.* (1994). doi:10.1101/gad.8.15.1829

24. Kirchhamer, C. V., Bogarad, L. D. & Davidson, E. H. Developmental expression of synthetic cis-regulatory systems composed of spatial control elements from two different genes. *Proc. Natl. Acad. Sci.* (1996). doi:10.1073/pnas.93.24.13849

25. Lander, E. S. The new genomics: Global views of biology. *Science* (1996). doi:10.1126/science.274.5287.536

26. Sequencing Consortium, C. elegans. Genome sequence of the nematode C. elegans: A platform for investigating biology. *Science* (1998). doi:10.1126/science.282.5396.2012

27. Dunham, I. *et al.* The DNA sequence of human chromosome 22. *Nature* (1999). doi:10.1038/990031

28. Shoemaker, D. D. *et al.* Experimental annotation of the human genome using microarray technology. *Nature* (2001). doi:10.1038/35057141

29. Kellis, M., Patterson, N., Endrizzi, M., Birren, B. & Lander, E. S. Sequencing and comparison of yeast species to identify genes and regulatory elements. *Nature* (2003). doi:10.1038/nature01644

30. Shashikant, C. S., Kim, C. B., Borbely, M. A., Wang, W. C. & Ruddle, F. H. Comparative studies on mammalian Hoxc8 early enhancer sequence reveal a baleen whale-specific deletion of a cis-acting element. *Proc. Natl. Acad. Sci. U. S. A.* (1998).

31. Sumiyama, K., Kim, C. B. & Ruddle, F. H. An efficient cis-element discovery method using multiple sequence comparisons based on evolutionary relationships. *Genomics* (2001). doi:10.1006/geno.2000.6422

32. Boffelli, D., Nobrega, M. A. & Rubin, E. M. Comparative genomics at the vertebrate extremes. *Nature Reviews Genetics* (2004). doi:10.1038/nrg1350

33. Alter, O., Brown, P. O. & Botstein, D. Singular value decomposition for genome-wide

expression data processing and modeling. *Proc. Natl. Acad. Sci. U.S.A.* **97**, 10101–10106 (2000).

34. Segal, E. *et al.* Module networks: identifying regulatory modules and their condition-specific regulators from gene expression data. *Nat. Genet.* **34**, 166–176 (2003).

35. DeRisi, J. *et al.* Use of a cDNA microarray to analyse gene expression patterns in human cancer. *Nat. Genet.* (1996). doi:10.1038/ng1296-457

36. Wang, K. *et al.* Monitoring gene expression profile changes in ovarian carcinomas using cDNA microarray. *Gene* (1999). doi:10.1016/S0378-1119(99)00035-9

37. Golub, T. R. *et al.* Molecular classification of cancer: Class discovery and class prediction by gene expression monitoring. *Science (80-. ).* (1999). doi:10.1126/science.286.5439.531

38. Wen, X. *et al.* Large-scale temporal gene expression mapping of central nervous system development. *Proc. ...* (1998). doi:10.1073/pnas.95.1.334

39. Ihmels, J. *et al.* Revealing modular organization in the yeast transcriptional network. *Nat. Genet.* (2002). doi:10.1038/ng941

40. Alon, U. *et al.* Broad patterns of gene expression revealed by clustering analysis of tumor and normal colon tissues probed by oligonucleotide arrays. *Proc. Natl. Acad. Sci. U. S. A.* (1999). doi:10.1073/pnas.96.12.6745

41. Wagner, A., Regev, A. & Yosef, N. Revealing the vectors of cellular identity with single-cell genomics. *Nature Biotechnology* (2016). doi:10.1038/nbt.3711

42. Shoemaker, D. D., Lashkari, D. A., Morris, D., Mittmann, M. & Davis, R. W. Quantitative phenotypic analysis of yeast deletion mutants using a highly parallel molecular bar-coding strategy. *Nat. Genet.* (1996). doi:10.1038/ng1296-450

43. Winzeler, E. A. *et al.* Functional characterization of the S. cerevisiae genome by gene deletion and parallel analysis. *Science (80-. ).* (1999). doi:10.1126/science.285.5429.901

44. Giaever, G. *et al.* Functional profiling of the Saccharomyces cerevisiae genome. *Nature* (2002). doi:10.1038/nature00935

45. Ideker, T. *et al.* Integrated genomic and proteomic analyses of a systematically perturbed metabolic network. *Science (80-. ).* (2001). doi:10.1126/science.292.5518.929

46. Fire, A. *et al.* Potent and specific genetic interference by double-stranded RNA in caenorhabditis elegans. *Nature* (1998). doi:10.1038/35888

47. Cong, L. *et al.* Multiplex genome engineering using CRISPR/Cas systems. *Science* **339**, 819–23 (2013).

48. Mali, P. *et al.* RNA-guided human genome engineering via Cas9. *Science (80-. ).* (2013). doi:10.1126/science.1232033

49. Hwang, W. Y. *et al.* Efficient genome editing in zebrafish using a CRISPR-Cas system. *Nat. Biotechnol.* (2013). doi:10.1038/nbt.2501

50. Boutros, M. *et al.* Genome-Wide RNAi Analysis of Growth and Viability in Drosophila Cells. *Science (80-. ).* (2004). doi:10.1126/science.1091266

51. Wang, T., Wei, J. J., Sabatini, D. M. & Lander, E. S. Genetic screens in human cells using

the CRISPR-Cas9 system. *Science* **343**, 80–4 (2014).

52. Parnas, O. *et al.* A Genome-wide CRISPR Screen in Primary Immune Cells to Dissect Regulatory Networks. *Cell* **162**, 675–686 (2015).

53. Park, R. J. *et al.* A genome-wide CRISPR screen identifies a restricted set of HIV host dependency factors. *Nat. Genet.* (2017). doi:10.1038/ng.3741

54. Tong, A. H. Y. *et al.* Global Mapping of the Yeast Genetic Interaction Network. *Science (80-. ).* (2004). doi:10.1126/science.1091317

55. Costanzo, M. *et al.* The Genetic Landscape of a Cell. *Science (80-. ).* **327**, 425–431 (2010).

56. Costanzo, M. *et al.* A global genetic interaction network maps a wiring diagram of cellular function. *Science (80-. ).* **353**, aaf1420–aaf1420 (2016).

57. Kuzmin, E. *et al.* Systematic analysis of complex genetic interactions. *Science (80-. ).* **360**, eaao1729 (2018).

58. Capaldi, A. P. *et al.* Structure and function of a transcriptional network activated by the MAPK Hog1. *Nat. Genet.* (2008). doi:10.1038/ng.235

59. Hughes, T. R. *et al.* Functional discovery via a compendium of expression profiles. *Cell* (2000). doi:10.1016/S0092-8674(00)00015-5

60. Fuchs, F. *et al.* Clustering phenotype populations by genome-wide RNAi and multiparametric imaging. *Mol. Syst. Biol.* (2010). doi:10.1038/msb.2010.25

61. Amit, I. *et al.* Unbiased Reconstruction of a Mammalian Transcriptional Network Mediating Pathogen Responses. *Science (80-. ).* **326**, 257–263 (2009).

62. Yosef, N. *et al.* Dynamic regulatory network controlling TH 17 cell differentiation. *Nature* (2013). doi:10.1038/nature11981

63. Adamson, B. *et al.* A Multiplexed Single-Cell CRISPR Screening Platform Enables Systematic Dissection of the Unfolded Protein Response. *Cell* **167**, 1867-1882.e21 (2016).

64. Datlinger, P. *et al.* Pooled CRISPR screening with single-cell transcriptome readout. *Nat. Methods* (2017). doi:10.1038/nmeth.4177

65. Norman, T. M. *et al.* Exploring genetic interaction manifolds constructed from rich single-cell phenotypes. *Science (80-. ).* (2019). doi:10.1126/science.aax4438

66. Groot, R., Lüthi, J., Lindsay, H., Holtackers, R. & Pelkmans, L. Large-scale image-based profiling of single-cell phenotypes in arrayed CRISPR-Cas9 gene perturbation screens. *Mol. Syst. Biol.* (2018). doi:10.15252/msb.20178064

67. Feldman, D. *et al.* Pooled optical screens in human cells. *bioRxiv* (2018). doi:10.1101/383943

68. Box, G. E. P. & Draper, N. R. Robust designs. *Biometrika* (1975). doi:10.1093/biomet/62.2.347

69. Box, G. E. P. Science and statistics. *J. Am. Stat. Assoc.* (1976). doi:10.1080/01621459.1976.10480949

70. Johnson, W. B. & Lindenstrauss, J. Extensions of Lipschitz mappings into a Hilbert space. *Contemp. Math.* **26**, 189–206 (1984).

71. Achlioptas, D. Database-friendly random projections. *Proc. Twent. ACM SIGMOD-SIGACT-SIGART Symp. Princ. database Syst.* 274–281 (2001). doi:10.1145/375551.375608

72. Dasgupta, S. & Gupta, A. An Elementary Proof of a Theorem of Johnson and Lindenstrauss. *Random Struct. Algorithms* **22**, 60–65 (2003).

73. Dasgupta, S. Experiments with Random Projection. in *Uncertainty in Artificial intelligence:Proceedings of 16th Conference* (2000).

74. Baraniuk, R. G. & Wakin, M. B. Random projections of smooth manifolds. *Found. Comput. Math.* (2009). doi:10.1007/s10208-007-9011-z

75. Rahimi, A. & Recht, B. Random Features for Large Scale Kernel Machines. *Adv. Neural Inf. Process. Syst.* (2007).

76. Rahimi, A. & Recht, B. Weighted sums of random kitchen sinks: Replacing minimization with randomization in learning. *Adv. neural Inf. Process. …* (2009).

77. Gilbert, A. C., Zhang, Y., Lee, K., Zhang, Y. & Lee, H. Towards understanding the invertibility of convolutional neural networks. in *IJCAI International Joint Conference on Artificial Intelligence* (2017).

78. Candes, E., Romberg, J. & Tao, T. Robust Uncertainty Principles : Exact Signal Reconstruction from Highly Incomplete Frequency Information. 1–41 (2005).

79. Donoho, D. L. Compressed sensing. *IEEE Trans. Inf. Theory* **52**, 1289–1306 (2006).

80. Candes, E. & Tao, T. Decoding by Linear Programming. **40698**, 1–22 (2004).

81. Candès, E. J. & Recht, B. Exact matrix completion via convex optimization. *Found. Comput. Math.* 1–49 (2009). doi:10.1007/s10208-009-9045-5

82. Gandy, S., Recht, B. & Yamada, I. Tensor completion and low-n-rank tensor recovery via convex optimization. *Inverse Probl.* (2011). doi:10.1088/0266-5611/27/2/025010

83. Donoho, D. L. & Tanner, J. Precise undersampling theorems. *Proc. IEEE* (2010). doi:10.1109/JPROC.2010.2045630

84. Kane, D. M. & Nelson, J. Sparser johnson-lindenstrauss transforms. *J. ACM* **61**, 4 (2014).

85. Rohban, M. H. *et al.* Systematic morphological profiling of human gene and allele function via cell painting. *Elife* (2017). doi:10.7554/eLife.24060

86. Barak, B. & Moitra, A. Noisy tensor completion via the sum-of-squares hierarchy. in *Conference on Learning Theory* 417–445 (2016).

87. Parvaresh, F., Vikalo, H., Misra, S. & Hassibi, B. Recovering sparse signals using sparse measurement matrices in compressed DNA microarrays. *IEEE J. Sel. Top. Signal Process.* (2008). doi:10.1109/JSTSP.2008.924384

88. Aghazadeh, A. *et al.* Universal microbial diagnostics using random DNA probes. *Sci. Adv.* (2016). doi:10.1126/sciadv.1600025

89. Cleary, B. *et al.* Detection of low-abundance bacterial strains in metagenomic datasets by eigengenome partitioning. *Nat. Biotechnol.* **33**, 1053–1060 (2015).

90. Alneberg, J. *et al.* Binning metagenomic contigs by coverage and composition. *Nat. Methods* **11**, (2014).

91. AlQuraishi, M. & McAdams, H. H. Direct inference of protein-DNA interactions using compressed sensing methods. *Proc. Natl. Acad. Sci. U. S. A.* (2011). doi:10.1073/pnas.1106460108

92. Golan, D., Erlich, Y. & Rosset, S. Weighted pooling-practical and cost-effective techniques for pooled high-throughput sequencing. *Bioinformatics* (2012). doi:10.1093/bioinformatics/bts208

93. de Boer, C. G. *et al.* Deciphering eukaryotic gene-regulatory logic with 100 million random promoters. *Nat. Biotechnol.* 1–10 (2019).

94. Bogard, N., Linder, J., Rosenberg, A. B. & Seelig, G. A Deep Neural Network for Predicting and Engineering Alternative Polyadenylation. *Cell* (2019). doi:10.1016/j.cell.2019.04.046

95. Bajikar, S. S., Fuchs, C., Roller, A., Theis, F. J. & Janes, K. A. Parameterizing cell-to-cell regulatory heterogeneities via stochastic transcriptional profiles. *Proc. Natl. Acad. Sci. U. S. A.* (2014). doi:10.1073/pnas.1311647111

96. Jaitin, D. A. *et al.* Massively parallel single-cell RNA-seq for marker-free decomposition of tissues into cell types. *Science* **343**, 776–9 (2014).

97. Shekhar, K. *et al.* Comprehensive Classification of Retinal Bipolar Neurons by Single-Cell Transcriptomics. *Cell* **166**, 1308-1323.e30 (2016).

98. Heimberg, G., Bhatnagar, R., El-Samad, H. & Thomson, M. Low Dimensionality in Gene Expression Data Enables the Accurate Extraction of Transcriptional Programs from Shallow Sequencing. *Cell Syst.* **2**, 239–250 (2016).

99. Shalek, A. K. *et al.* Single-cell transcriptomics reveals bimodality in expression and splicing in immune cells. *Nature* **498**, 236–40 (2013).

100. Bates, M., Huang, B., Dempsey, G. T. & Zhuang, X. Multicolor super-resolution imaging with photo-switchable fluorescent probes. *Science (80-. ).* (2007). doi:10.1126/science.1146598

101. Berger, B., Daniels, N. M. & William Yu, Y. Computational biology in the 21st century: Scaling with compressive algorithms. *Communications of the ACM* (2016). doi:10.1145/2957324

102. Cleary, B., Cong, L., Cheung, A., Lander, E. S. & Regev, A. Efficient Generation of Transcriptomic Profiles by Random Composite Measurements. *Cell* **171**, 1424-1436.e18 (2017).

103. Cleary, B. *et al.* Compressed sensing for imaging transcriptomics. *bioRxiv* 743039 (2019).

104. Chen, K. H., Boettiger, A. N., Moffitt, J. R., Wang, S. & Zhuang, X. Spatially resolved, highly multiplexed RNA profiling in single cells. *Science (80-. ).* **348**, aaa6090–aaa6090 (2015).

105. Angelo, M. *et al.* Multiplexed ion beam imaging of human breast tumors. *Nat. Med.* **20**, 436–42 (2014).

106. Goltsev, Y. *et al.* Deep Profiling of Mouse Splenic Architecture with CODEX Multiplexed Imaging. *Cell* 203166 (2018). doi:10.1016/j.cell.2018.07.010

107. Saka, S. K. *et al.* Highly multiplexed in situ protein imaging with signal amplification by. *bioRxiv* (2018). doi:10.1101/507566

108. Hu, F. *et al.* Supermultiplexed optical imaging and barcoding with engineered polyynes. *Nat. Methods* (2018). doi:10.1038/nmeth.4578

109. Kishi, J. Y. *et al.* SABER amplifies FISH: enhanced multiplexed imaging of RNA and DNA in cells and tissues. *Nat. Methods* (2019). doi:10.1038/s41592-019-0404-0

110. Boufounos, P. T. & Baraniuk, R. G. 1-Bit compressive sensing. in *CISS 2008, The 42nd Annual Conference on Information Sciences and Systems* 16–21 (2008). doi:10.1109/CISS.2008.4558487

111. Plan, Y. & Vershynin, R. Robust 1-bit compressed sensing and sparse logistic regression: A convex programming approach. *IEEE Trans. Inf. Theory* **59**, 482–494 (2013).

112. Gasperini, M. *et al.* A Genome-wide Framework for Mapping Gene Regulation via Cellular Genetic Screens. *Cell* (2019). doi:10.1016/j.cell.2018.11.029

113. Campa, C. C., Weisbach, N. R., Santinha, A. J., Incarnato, D. & Platt, R. J. Multiplexed genome engineering by Cas12a and CRISPR arrays encoded on single transcripts. *Nat. Methods* (2019). doi:10.1038/s41592-019-0508-6

114. Macosko, E. Z. *et al.* Highly Parallel Genome-wide Expression Profiling of Individual Cells Using Nanoliter Droplets. *Cell* **161**, 1202–1214 (2015).

115. Cao, J. *et al.* The single-cell transcriptional landscape of mammalian organogenesis. *Nature* (2019). doi:10.1038/s41586-019-0969-x

116. Kehe, J. *et al.* Massively parallel screening of synthetic microbial communities. *Proc. Natl. Acad. Sci. U. S. A.* (2019). doi:10.1073/pnas.1900102116

117. Jeanne, J. M., Fişek, M. & Wilson, R. I. The Organization of Projections from Olfactory Glomeruli onto Higher-Order Neurons. *Neuron* (2018). doi:10.1016/j.neuron.2018.05.011

118. Liu, X., Li, Y. I. & Pritchard, J. K. Trans effects on gene expression can drive omnigenic inheritance. *bioRxiv* (2018). doi:10.1101/425108

119. Wray, N. R., Wijmenga, C., Sullivan, P. F., Yang, J. & Visscher, P. M. Common Disease Is More Complex Than Implied by the Core Gene Omnigenic Model. *Cell* (2018). doi:10.1016/j.cell.2018.05.051

120. Udler, M. S. *et al.* Type 2 diabetes genetic loci informed by multi-trait associations point to disease mechanisms and subtypes: A soft clustering analysis. *PLoS Med.* (2018). doi:10.1371/journal.pmed.1002654

121. Smillie, C. S. *et al.* Intra- and Inter-cellular Rewiring of the Human Colon during Ulcerative Colitis. *Cell* (2019). doi:10.1016/j.cell.2019.06.029

122. Genomes Project Consortium, 1000. An integrated map of genetic variation from 1,092

human genomes. *Nature* (2012). doi:10.1038/nature11632

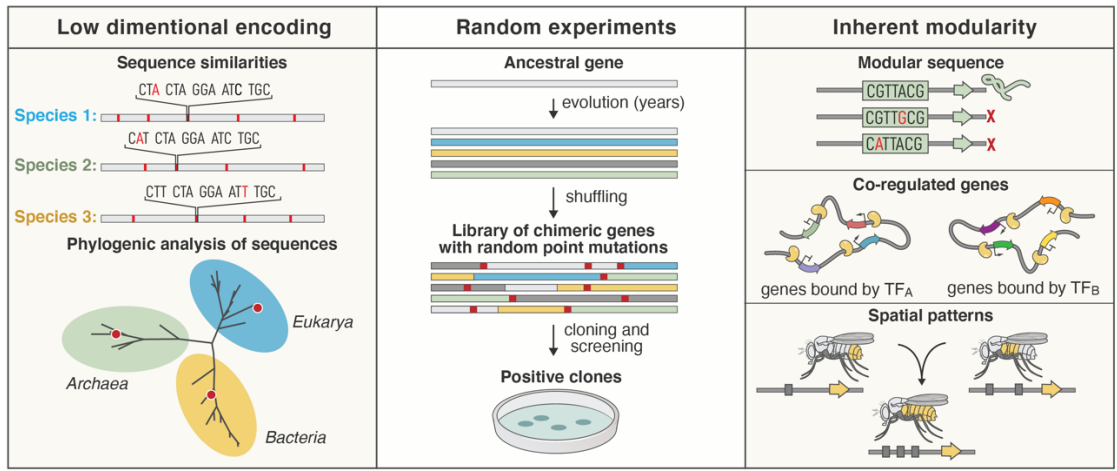

**Figure 1. Low-dimensionality, randomness, and modularity in biological experiments and analysis**

High-dimensional biological systems can be efficiently studied through seemingly under-sampled experiments. In phylogenetic analysis (**left**), though genomes of many species are large and the organisms themselves complex, relatively small genomic regions can encode the evolutionary relatedness of a large number of species. In DNA shuffling (**middle**), large libraries of candidate genes to be screened for a desired phenotype can be quickly constructed through a random cloning process that generates both chimeric sequences and point mutations. The study of gene regulatory networks (**right**) is dramatically simplified by inherent modularity in *cis*-regulatory elements (**top**), co-regulated sets of genes (**middle**), and coherent spatial patterns (**bottom**).

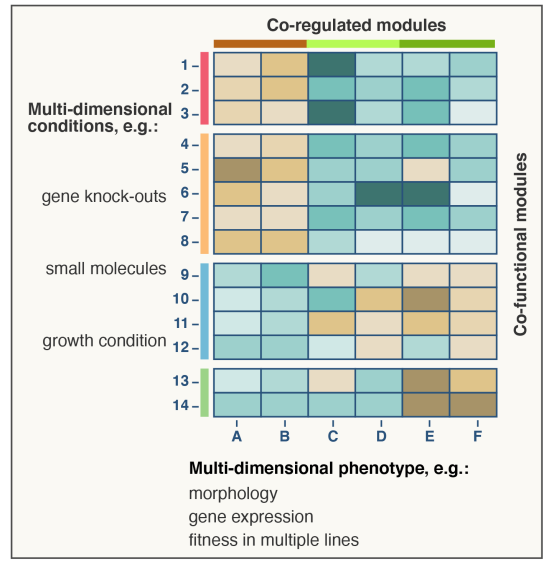

**Figure 2. Co-regulated and co-functional modules**

Multi-dimensional phenotypes (**columns of matrix**) can often be organized into co-regulated modules; for example, levels (blue/brown) of gene expression or morphological features (columns) that co-vary across samples (rows). Different perturbations or conditions (**rows**) can be grouped into co-functional modules; for example, sets of knock-outs or small molecules that produce similar phenotypes (columns).

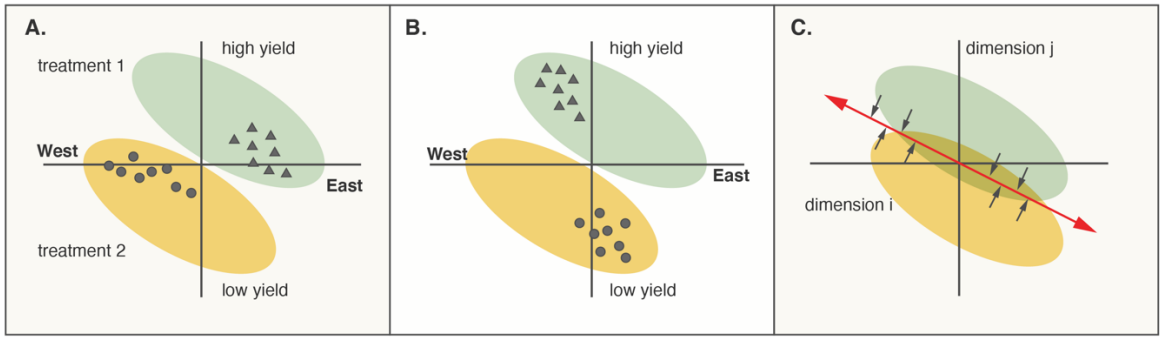

**Figure 3. Amplification or suppression of differences between populations**

Technical or confounding factors can manifest in directions of significant variation in data. Experimental sampling (**A** and **B**) or computational analysis (**C**) that is aligned with these directions can result in skewed results. (**A** and **B**) Crop yield can be affected by both East-West location of samples (triangles and circles) in a field (a confounder), and by treatment. Ellipses depict hypothetical yield distributions (y axis) for each treatment (green, yellow), as a function of location (x axis). If all samples for each treatment are grouped at opposite ends of the field, then the observed differences in yield will either be suppressed (**A**) or amplified (**B**). (**C**) Two populations have substantial differences in dimension j (vertical axis). When the data are projected onto the first principal component (red line; grey arrows depict projection onto the line), the difference in the distributions is nearly eliminated.

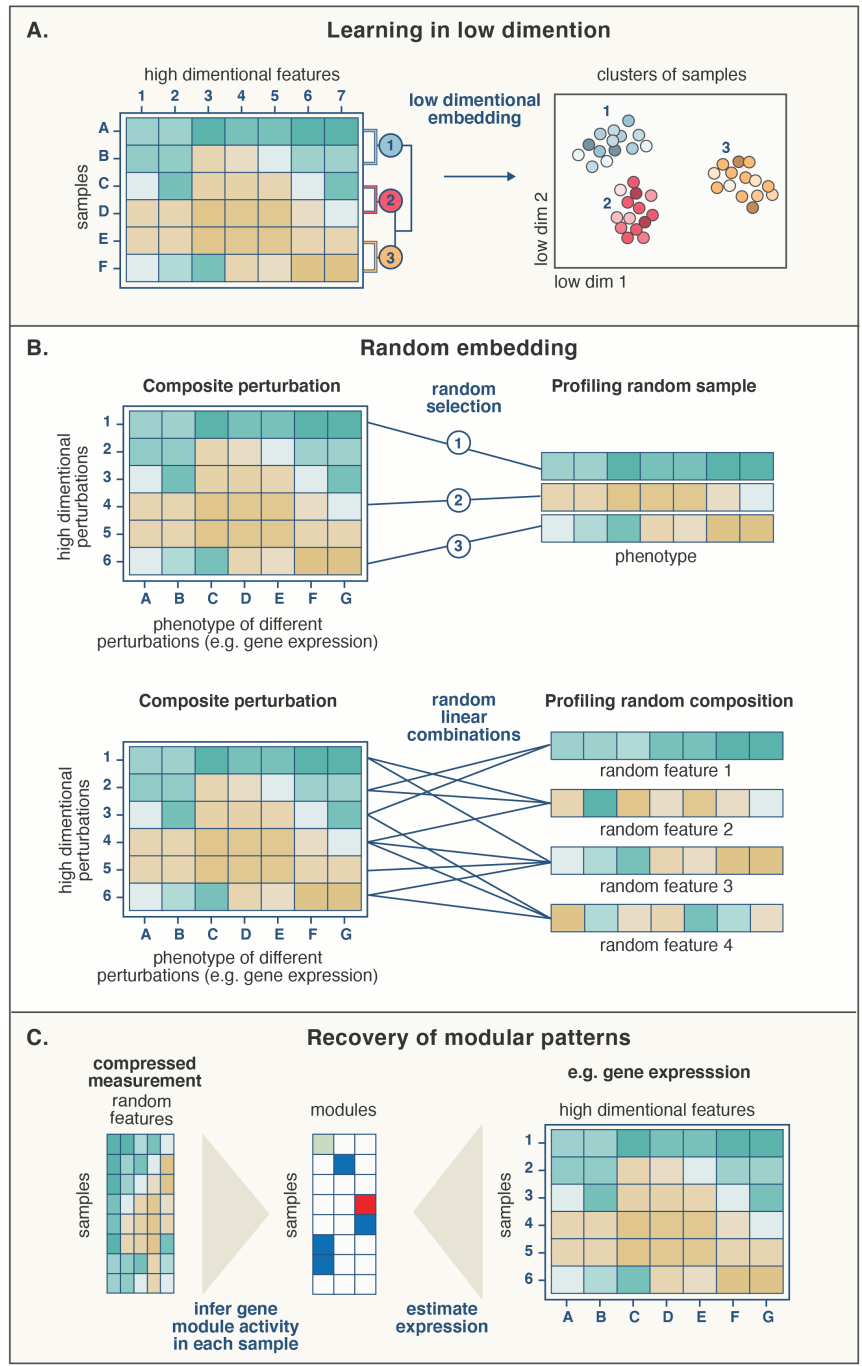

**Figure 4: Conceptual framework for random and composite experiments**

(**A**) Learning in low dimensions. The similarity of samples in a high-dimensional space (rows in matrix, left) can be preserved in a low-dimensional embedding (dots, right). (**B**) Random embedding. Low-dimensional embeddings can be defined by random features, corresponding, for instance, to random samples (**top**) or random linear combinations of high-dimensional features (**bottom**). (**C**) Recovery of modular patterns. In compressed sensing, relatively few random features in each sample (**left**) are used to infer sparse, latent parameters (**middle**; *e.g.*, sparse gene module activities), from which high-dimensional features can be recovered (**right**; *e.g.*, individual gene expression levels).

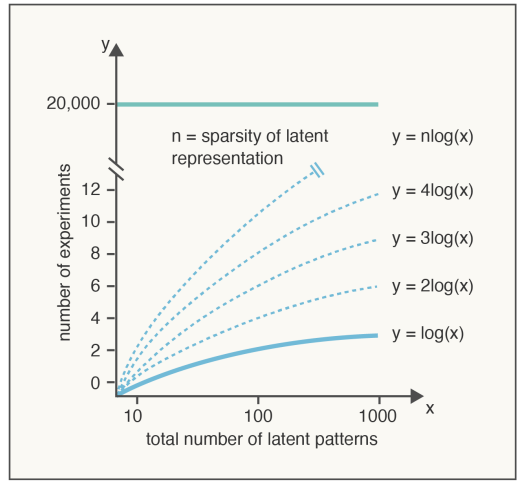

**Figure 5: Approximate scaling of compressed experiments**

When high-dimensional vectors can be accurately represented by a combination of few (up to *n*) latent patterns, compressed experiments can be nearly exponentially more efficient. The number of compressed experiments (*y axis*) scales linearly with n, and logarithmically with the total number of patterns that could have been combined (**x axis**). If biological structure is ignored, variables must be tested individually, for example by measuring or perturbing each gene separately (horizontal line, y=20,000).